\documentclass[
aps,
prx,prl,nc
amsmath,amssymb,
reprint,
superscriptaddress,
amsart,
]{revtex4-2}

\usepackage{graphicx}
\usepackage[utf8]{inputenc}
\graphicspath{ {images/} }
\usepackage{units}
\usepackage{amsmath}
\usepackage{ulem}
\usepackage{color}
\usepackage{lineno}

\usepackage{fancyhdr}
\begin{document}

\title{Metagrating-based Single-pixel Acoustic Direction Finding}


\author{Thomas Macleod}
\email{t.macleod@student.unsw.edu.au} 
\affiliation{%
School of Engineering and Technology, University of New South Wales, Canberra, Australia
}

\author{Sebastian Oberst}
\altaffiliation[Also at ]{School of Engineering and Technology, University of New South Wales, Canberra, Australia}
\affiliation{%
Centre for Audio, Acoustics and Vibration, University of Technology Sydney, Sydney, Australia
}%



\author{David A.~Powell}
\affiliation{%
School of Engineering and Technology, University of New South Wales, Canberra, Australia
}%

\author{Yan Kei Chiang}
\email{y.chiang@unsw.edu.au} 
\affiliation{%
School of Engineering and Technology, University of New South Wales, Canberra, Australia
}%

\date{\today}

\begin{abstract}
Acoustic metamaterials provide new opportunities for compact and efficient wavefront manipulation, extending beyond conventional bulky and power-intensive phased-arrays. In this work, we exploit the spatial encoding properties of the acoustic metagrating aperture to transform incident acoustic fields into compressed measurements for single-pixel acoustic source localisation. The proposed method enables accurate direction finding of acoustic sources over both 180 and 360 degrees angular ranges. Numerical simulations confirm the robustness of the metagrating-based compressive sensing approach against noise and limited sampling. Experimental validation is conducted to verify its feasibility with practical metagrating prototyes. Compared wiith tranditional array-based localisation techniques, the single-pixel metagrating system significantly reduces hardward complexity while maintaining high localisation accuracy. These findings demonstrate the potential of integrating compressive sensing with acoustic metagratings for compact, low-cost, scalable source detection systems, with prospective applications in industrial monitoring, target tracking and non-destructive health monitoring.

\end{abstract}

\maketitle

Acoustic source localisation and direction-finding have received significant attention over recent decades due to its wide variety of potential applications ranging from fault detection to medical sensing \cite{Erol_underwaterlocalise_2011, Siddiqui_railway_2022, Shin_medical_2017}.  Traditional direction-of-arrival (DOA) estimation techniques rely on microphone arrays that measure the time delay between the sound received at one sensor (microphone) location to another \cite{book_tradition_2022, Eranti_tradition_2022, Thomas_tradition_2006, Molaei_tradition_2024}. This time delay stems from the difference in path length the soundwave must travel between sensors, which when expressed in the frequency domains appears as a shift in the phase of the incoming signal. To estimate the DoA, various signal processing algorithms can be utilised, such as time-difference-of-arrival (TDOA), steered response power beamforming and high-resolution subspace methods. However, these methods rely on dense, synchronised arrays. High angular resolution requires a large number of elements arranged with sub-wavelength spacing and wide apertures which lead to high power consumption, increased hardware complexity, high data acquisition rates and complex calibration of each sensor within the array. The physical and practical constraints associated with these systems impose fundamental limitations on their scalability and deployment flexibility.

\begin{figure*}[hbtp!]
	\includegraphics[width=1\textwidth]{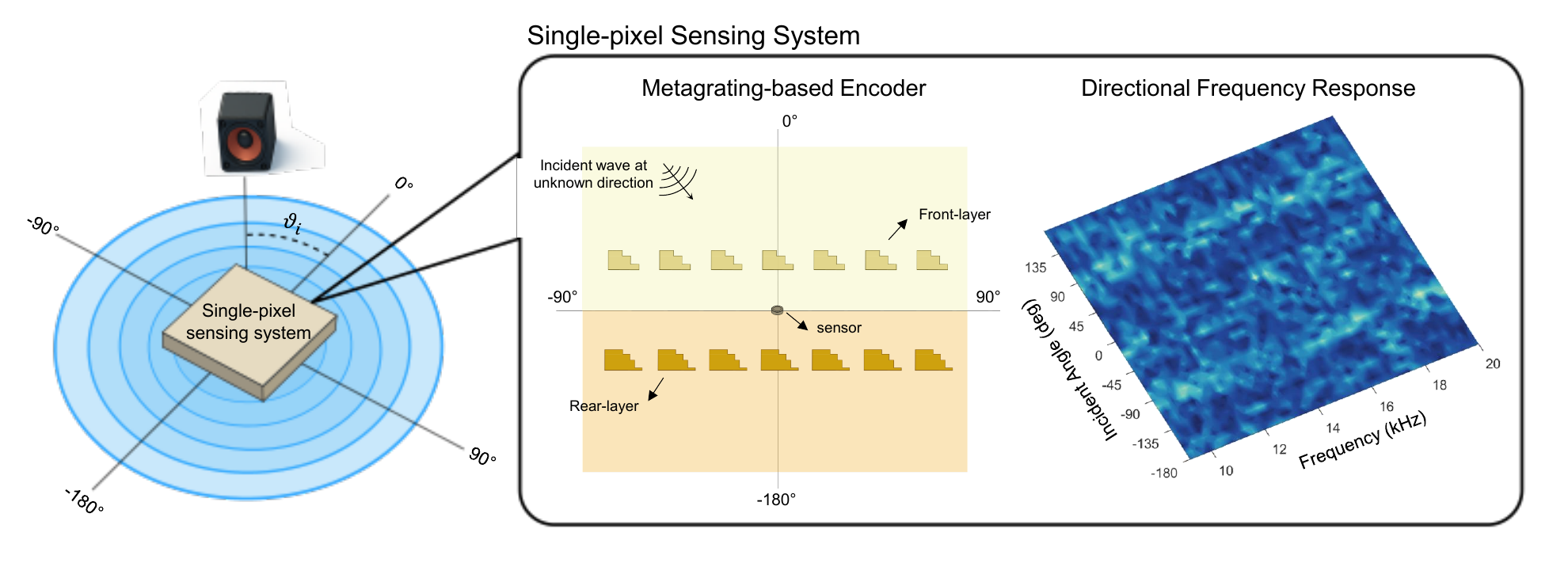}
	\caption{
	    Schematic diagram of the metagrating-based single-sensing system for source direction finding.
	    \label{fig:schematic}}
\end{figure*}

To overcome these limitations, researchers have explored the use of compressive sensing as a new framework for acoustic source localisation \cite{Gerstoft_cs_2018, Marzik_cs_2021}. Instead of sampling the sound field at or above the Nyquist rate using multiple microphones, compressive sensing technique reconstructs the signal by solving a sparse recovery problem from a small number of measurements. This approach has been applied to acoustic imaging, beamforming and holography, enabling acoustic field reconstruction with the use of a small subset of microphones. Recently, the integration of acoustic metamaterials with compressive sensing has opened new pathways for achieving DOA estimation through wave manipulation rather than sensor distribution. Acoustic metamaterials possess tailored dispersion, impedance ad phase characteristics that allow unprecedented control over wave propagation \cite{Chen_coiled_2018, Zhao_helical_2018, Li_helical_2020, Zabihi_am_2023, Le_Origami_2025, Yang_hr_2025}. Their ability to impose spatially varying phase responses enables them to transform an incident acoustic field into a set of uniquely modulated fields that contain spatial information. By combining with compressive sensing, the modulated signals can be used to reconstruct the direction of sound sources using only a single sensor. Researchers demonstrated the single-detector acoustic camera system based on subwavelength metamaterials, which provide the direction-dependent frequency response of the microphone, for acoustic source localization \cite{Xie_hrcs_2015, Jiang_coilingcs_2019, Sun_metacs_2020, wang_tccs_2025}. However, most reported metamaterial structures have complicated interior features and rely on resonant unit cells. The efficiency and resolution of this approach depend on the level of discretisation, meaning that dense arrays of complex, fine geometries are required that are not only difficult to fabricate, but are also difficult to scale for high frequencies as the geometries are too narrow.

In this study, we propose a metagrating-based compressive sensing framework for single-pixel acoustic direction finding. Unlike labyrinthine or Helmholtz-resonator-type metamaterials, metagratings manipulate diffraction orders rather than relying on local resonances. Emerging results on acoustic metagratings show that properly designed periodicity and metaatom geometry can realise robust, large-angle beam control with high efficiency \cite{Hou_grating_2019, Craig_grating_2019, Chiang_grating_2020}. Combining the metagrating design with the compressive sensing technique, it can provide high-efficiency wide-angle coding with minimal energy loss and naturally low-correlation measurement patterns via diffraction order selection and metagrating parameter tuning, generating effective sensing matrix for compressive reconstruction. In this work, we adopt the metagrating with periodically arranged identical metaatoms to encode incident pressure field. We pair this metagrating-based encoder with Orthogonal Matching Pursuit (OMP) to retrieve source angular direction from a single sensor. Numerical simulations have been conducted to demonstrate the source direction finding performance of our proposed system over 360-degree using single-pixel receiver. Noise effect and frequency bandwidth requirement are also numerically investigated. Experimental study has validated that this metagrating-based system can accurately estimate direction of a source over 360-degree angular range using only a single sensor. This integration of metagrating encoding with compressive sensing establishes a new pathway for compact, low-cost, and scalable acoustic localisation platforms.

\section{Theory and Design}
\subsection{Two-layer Metagrating Configuration}
Metagratings represent a new generation of wavefront manipulating structures that exploit engineered diffraction rather than subwavelength resonance  to contorl wave propagation. Unlike conventional gradient metasurfaces which depends on locally resonant elements, metagratings employ periodic arrangements of subwavelength scatterers to distribute acoustic energy among selected diffraction orders with high efficiency based on their bianisotropic properties. In this study, we build upon these characteristics, using metagrating as a spatial encoder for compressive sensing. When an acoustic wave impinges on the metagrating, the structure transforms the incident field into a set of direction-dependent diffracted wavefronts. Each incident angle produces a unique scattering "signature", which serves as a distinct measurement mode. This provides a sensing matrix that relates the incoming acoustic field to the modulated pressure received by a single sensor positioned behind the metagrating as illustrated in Fig~\ref{fig:schematic}.

\begin{figure}[hbtp!]
	\includegraphics[width=0.45\textwidth]{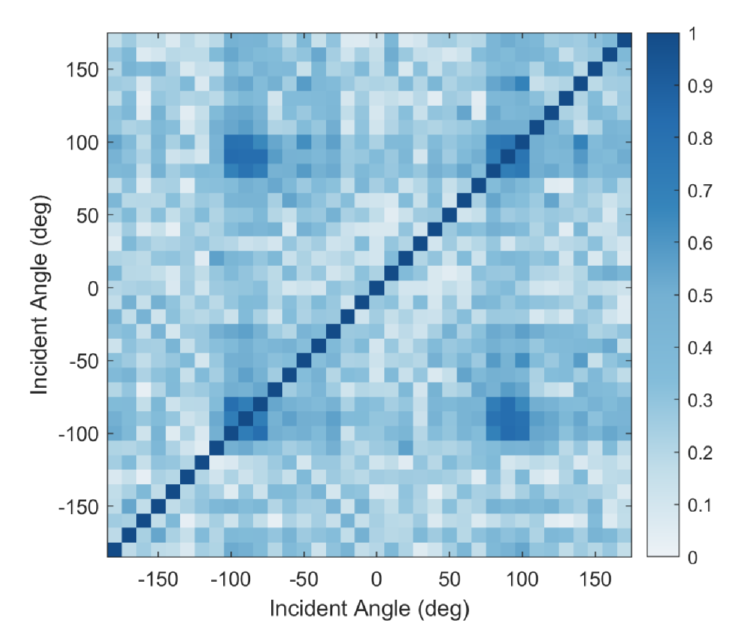}
	\caption{
	    Correlation of the proposed two-layer finite metagrating-encoder under monopole source excitation.
	    \label{fig:corr}}
\end{figure}

To ensure the encoding capability across the full azimuthal range, a two-layer metagrating configuration is proposed. Each layer is designed to encode the incident field within a $180^\circ$ sector, and the combination of the two enables complete $360^\circ$ direction estimation. Each metagrating layer consists of periodical array of multilayered stepped metaatoms. The periodicity $L$ of the metagrating is selected such that multiple propagating diffraction orders exist within the $180^\circ$ range based on Bragg's condition
\begin{equation}\label{eq:mtheta}
    \sin{\theta_{m}} = \sin{\theta_{i}} + m\lambda / L,
\end{equation}
where $\theta_{i}$ and $\theta_{m}$ are the angles of incidence and diffraction for $m$-th order, respectively, and $\lambda$ is the acoustic wavelength at frequency $f$. Rather than maximising efficiency for a single diffraction order, the metaatoms are designed to minimise the correlation between the pressure responses corresponding to different incident angles. Each layer acts as a spatial encoder for its respective hemisphere. The two layers are separated by an air gap of $d$ = 20 mm, and the sensor is centrally located behind both layers at $(x,y)=(0,0)$ to capture the composite pressure field as shown Fig~\ref{fig:schematic}.

\subsection{Correlation Analysis}
Numerical simulations are performed with COMSOL Multiphysics 6.2 for evaluating the total field distribution with respect to incident direction over $360^\circ$ in $10^\circ$ increments. Figure~\ref{fig:schematic} demonstrates the magnitude of the total pressure field received by the single sensor across the frequency range from 9.2kHz to 20kHz, showing distinct angular response generated by the two-layer metagrating for different angles of incidence. To evaluate the angular encoding performance of the metagrating, the correlation $\rho(\theta_{i,j},\theta_{i,k})$ between two incident angles $\theta_{i,j}$ and $\theta_{i,k}$ is evaluated through multi-frequency correlation analysis
\begin{equation}\label{eq:correlateeq}
    \rho(\theta_{i,j},\theta_{i,k}) = \frac{|\langle \textbf{h}(\theta_{i,j}),\textbf{h}(\theta_{i,k}) \rangle |}{\parallel \textbf{h}(\theta_{i,j}) \parallel_2 \parallel \textbf{h}(\theta_{i,k}) \parallel_2}, \text{with } j,k=1,...,N
\end{equation}
where $\langle \cdot,\cdot \rangle$ denotes the complex inner product, $\parallel \cdot \parallel_2$ is the Eclidean norm, $\textbf{h}(\theta)$ is the complex total pressure vector obtained at the sensor position, i.e., $\textbf{h}(\theta) = [p(\theta, f_1), \dots, p(\theta, f_K)]^\text{T}$, $j$ and $k$ are the indices of the testing incident angles, and $N$ denotes the total number of possible source angles. A smaller correlation coefficient indicates that the total pressure responses for these two angles are highly distinct. The full correlation matrix of the proposed two-layer finite metagrating-encoder formed by 15 metaatoms in each layer under monopole source excitation is shown in Fig~\ref{fig:corr}. It exhibits a dark diagonal (self-correlation) and light off-diagonal regions, indicating highly distinct scattering responses and significant angular separability. Its geometric parameters are listed in Table I in Supplementary Note 1.

Compared with labyrinthine or resonant metmaterials that rely on deep cavities or coiling channels, the present metagrating-based approach encodes angular information through interference among multiple diffraction orders. The coarse discretization of metagrating minimises thermo-viscous losses, enabling higher acoustic efficiency and stable performance over a wide frequency band. In addition, the scattering behavior of the metagrating depends primarily on the geometric ratio of structural parameters to wavelength. Therefore, the same encoding principle can be extended to different frequency bands without redesigning the metaatoms.

\begin{figure}[hbtp!]
	\includegraphics[width=0.45\textwidth]{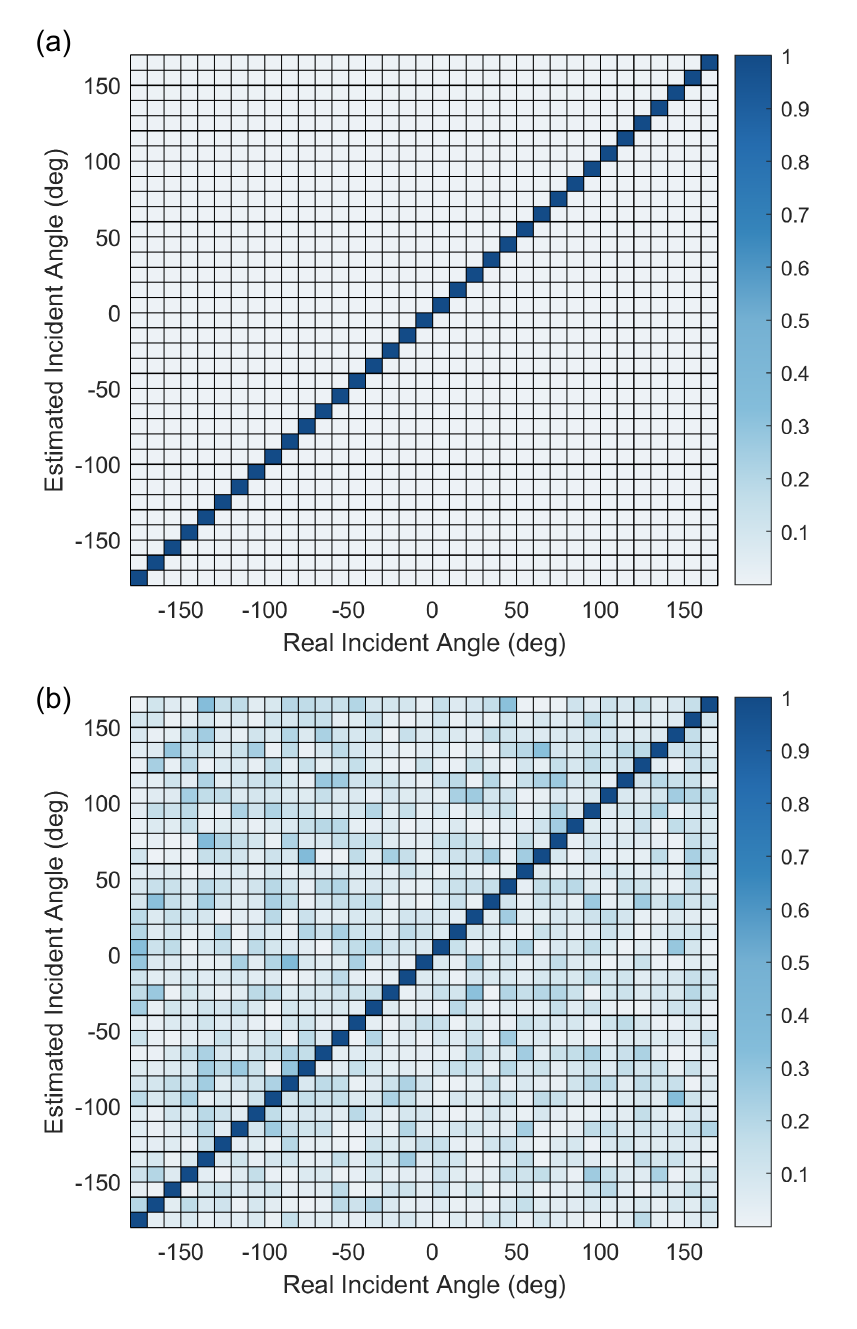}
	\caption{
	    \textbf{Numerical results of a single source localization.}
        (a) Without noise. (b) With Gaussian noise at SNR of 15dB.
	    \label{fig:DOA}}
\end{figure}
\begin{figure*}[hbtp]
	\includegraphics[width=0.85\textwidth]{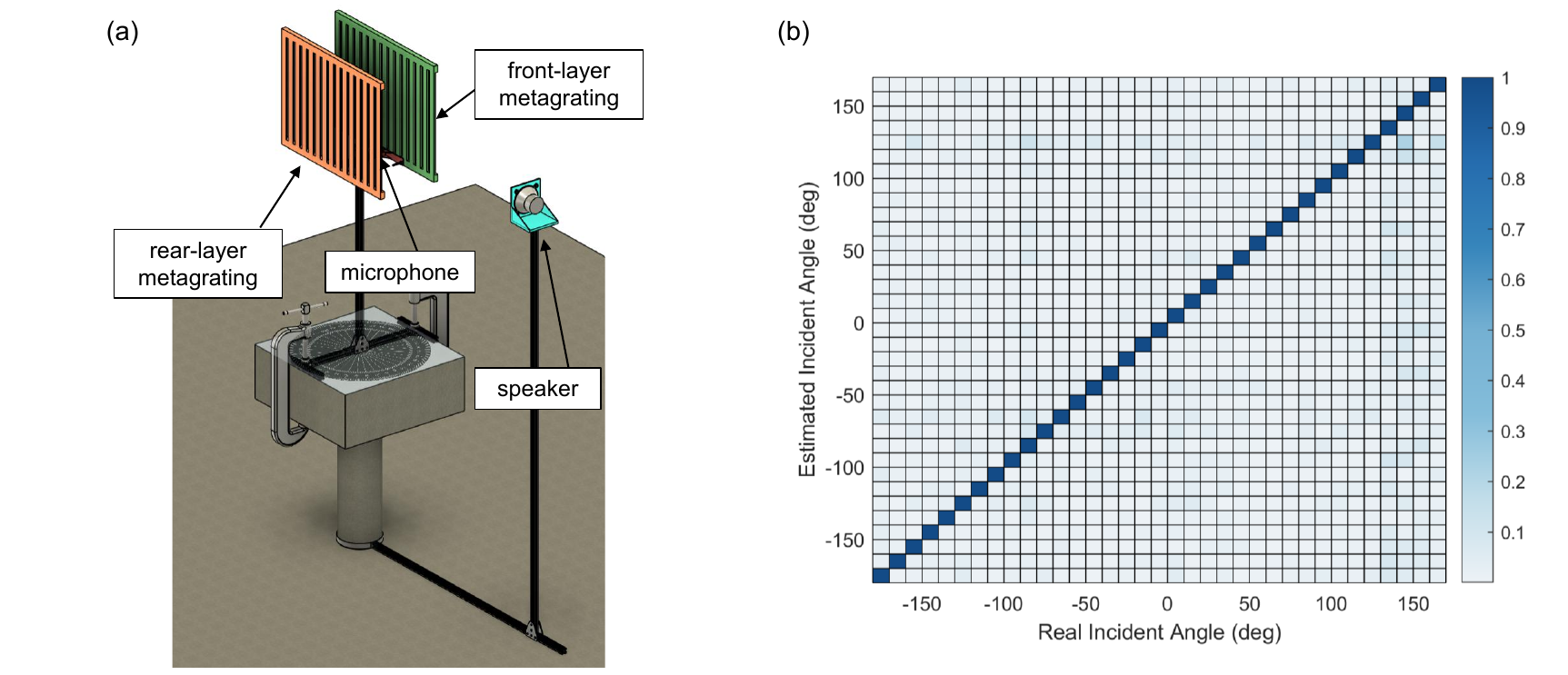}
	\caption{
	    \textbf{Experimental demonstration of single-pixel source direction finding.}
	    (a) Schematic experimental setup. (b)
        Experimental source angular direction finding performance across the full azimuthal range.
	    \label{fig:exp}}
\end{figure*}

\subsection{Compressive Sensing Framework}
The two-layer metagrating system and a single sensor together form a physical implementation of the linear compressive sensing model \cite{Gerstoft_cs_2018}
\begin{equation}\label{eq:linearcs}
    \mathbf{y} = \Phi \mathbf{x} + \mathbf{n},
\end{equation}
where $\mathbf{y}$ represents a vector of the measured single-pixel output, $\mathbf{x}$ is the sparse source distribution vector, $\mathbf{n}$ denotes measurement noise, and $\Phi$ is known as the sensing matrix constructed using the numerically obtained signal-pixel responses $p(\theta_{i})$ for discrete incident angles between $-180^\circ$ and $170^\circ$. Each response was normalised by the incident pressure amplitude $|p_{i}|$. The columns of $\Phi$ exhibit low pairwise correlation due to the intentionally diverse angular scattering patterns, enhancing the stability and accuracy of sparse recovery algorithms. In this work, we use OMP, as outline in Ref.\cite{Khosravy_OMP_2020}, to iteratively identify the most significant source directions by selecting columns of $\Phi$ that are most correlated with measurement vector $\mathbf{y}$, and refine the estimate by orthogonally projecting the residual onto the subspace spanned by the selected elements of $\Phi$. This process continues until the residual error falls below a predefined threshold. The output is a sparse vector $\hat{\mathbf{x}}$, in which the nonzero entries correspond to the reconstructed source directions and their relative amplitudes.

\section{Source Direction Finding}

Numerical simulations are conducted to evaluate the source direction reconstruction performance of the proposed metagrating-based single-pixel system (see Section Methods). A monopole source is used to generate incident pressure. It is positioned on a circular arc of radius $r_s=1$m from the sensor position. The source angular position is varied in $10^\circ$ increments over the azimuthal range of $0^\circ$ to $360^\circ$ to obtain the total pressure field responses used for sensing matrix construction. The reconstructured source directions are obtained by applying the OMP algorithm to simulated pressure magnitude using the sensing matrix $\Phi$ described in Section I.C. Figure 3(a) shows the estimated single monopole source direction without noise using the two-layer metagrating configuration. The incident direction corresponding to the max reconstructed amplitude in $\hat{\mathbf{x}}$ is identified as the estimated source direction. The clear dark diagonal demonstrates that the proposed approach can precisely reconstruct the source direction with unity reconstruction strength at the correct source direction due to the sufficient encoding diversity provided by the metagrating. We have also performed a frequency limitation study to investigate the effects of the measurement frequency bandwidth on the overall reconstruction performance of the metagrating-based sensing system. Results in the Supplemantary Note 2 demonstrate that the proposed method can still perform good source direction estimation using only 5.2 kHz bandwidth with the second highest strength less than 0.44.

To assess the robustness against measurement noise, additional Gaussian noise at a target signal-to-noise ratio (SNR) is introduced into the simulated detected pressure field before reconstruction. With the SNR of 15 dB, the reconstructed directions remained accurate, showing dominate diagonal in Figure 3(b). The averaged second highest reconstructed strength across 100 Monte Carlo trials at an SNR of 15dB and 18dB is less than 0.31 and 0.23, respectively, across the entire $360^\circ$ sensing range (see Supplementary Note 3).


\section{Experimental Verification}
To validate our single-pixel sensing framework, the single loudspeaker direction estimation of our metagrating-based encoder is experimentally demonstrated. Our experiments are conducted inside an anechoic chamber with the setup shown in Fig.~\ref{fig:exp} (a). A two-layer metagrating was 3D printed using PLA filament to implement the $360^\circ$ spatial encoder described in Section I. The sensing matrix for experiments was built from single-point sensor responses measured while sweeping the source angle. We scanned $\theta_i \in [-180^\circ, 170^\circ]$ with a step of $10^\circ$. To remove system-level gain, a reference pressure amplitude $|p_{\text{ref}}|$ was acquired with the metagrating removed while keeping the source at $0^\circ$, detector and stage geometry unchanged. The metagrating-encoded measurements $|p_{\text{mea}}(\theta_i)|$ were then normalised by the reference data to compensate for the loudspeaker radiation pattern. The normalised directional frequency responses of the metagrating-encoder are used to construct the measurement sensing matrix $\Phi_{\text{mea}}$. By applying OMP to the measured spectra at the microphone, loudspeaker direction estimation was performed. As shown in Fig.~\ref{fig:exp} (b), the metagrating acoustic camera system can accurately localise
a sound source across a full 360° sensing range, with the second highest reconstruction level of 0.22.

\section{Conclusion}
We proposed a compact and scalable single-pixel acoustic direction finding system by using two-layer of metagratings. Unlike conventional micorphone arrays that require spatially distributed sensors, the proposed approach exploits the diffraction-based encoding created by the metagratings to transform spatial information into unique spectral signatures detectable by a single sensor. Numerical simulations demonstrated that the metagratings exhibit low inter-angle correlation over a wide frequency range, which ensure the system can effectively converts spectral signals into sparse angular reconstructions. This enables robust source localization across a full $360^\circ$ range with unity reconstruction strength at the correct source direction without noise. Furthermore, numerical results showed that the system is able to maintain the direction finding performance with the added Gaussian noise, i.e., the average second highest reconstructed strength is less than 0.31, at SNR of 15 dB. We have examined the effect of the frequency bandwidth on the source direction localization numerically by applying the top-down (reducing the upper frequency limit) and bottom-up restrictions (increasing the lower frequency limit). Accurate source direction estimated can still be achieved by using only 5.2 kHz bandwidth for both restrictions, i.e., the average second highest reconstructed strength is less than 0.44. We developed and fabricated the two-layer metagrating prototype to experimentally verified its single-pixel sensing performance. Results demonstrated that accurate direction of a loudspeaker can be achieved, with the second highest reconstruction level of 0.22.
In this study, we establish metagrating-based compressive sensing as a powerful paradigm for single-pixel acoustic direction finding and imaging. Its scalability, structural simplicity and robustness against noise may facilitate the further development of advanced acoustic systems for spatial sound mapping, robotic auditory perception, gesture recognition and damage detection.

\section{Methods}
\subsection{Numerical Simulations} \label{sec:methods_num}
To investigate the directional sensing performance of the proposed single-pixel metagrating system , 2D full-wave numerical simulations are performed using the Pressure Acoustics module of COMSOL Multiphysics 6.2. The computational domain consisted of an air region of density $\rho_0=1.24 $kg/m$^3$ and sound speed of $c_0=343 $m/s bounded by perfectly matched layers to avoid reflections. The metagratings was modeled as acoustically rigid structures. The maximum element size is set to $\lambda$/10.

\subsection{Acoustic Pressure Measurements} \label{sec:methods_exp}
The acoustic pressure measurements were conducted in an anechoic chamber with the experimental setup illustrated in Fig.~\ref{fig:exp} (a). The loudspeaker was driven by a tonal sine sweeping from 9.2kHz to 20kHz with an interval of 200Hz. The field measurements were obtained using a 1/4 in.~microphone (B\&K type 4135) connected to a microphone power supply (B\&K type 2807). National Instruements data acquisition board (NI-USB6251) was used to recorded the voltage output from the microphone.


\newpage
\bibliography{reference}


\end{document}